\documentclass[a4paper,11pt]{article}

\usepackage{jheppub}
\usepackage{lineno}
\usepackage{graphicx}
\usepackage{amsmath}
\usepackage{amssymb}
\usepackage{hyperref}
\usepackage{braket}
\usepackage{slashed}
\usepackage{comment}

\def\beq{\begin{equation}}
\def\eeq{\end{equation}}
\def\bc{\begin{cases}}
\def\ec{\end{cases}}
\def\bal{\begin{aligned}}
\def\eal{\end{aligned}}
\newcommand{\mc}{\mathcal}
\newcommand{\mf}{\mathfrak}

\arxivnumber{}
\title{\boldmath Schwinger Effect of Extremal Reissner-Nordstr\"{o}m Black Holes}
\author{Puxin Lin, Gary Shiu}
\affiliation{Department of Physics, University of Wisconsin-Madison\\
1150 University Avenue, Madison, WI 53706, USA\\}

\emailAdd{plin73@wisc.edu, shiu@physics.wisc.edu}

\abstract{The Schwinger effect has a variety of physics applications. In the context of black hole physics, it provides a channel for the decay of charged black holes. While the Schwinger rate has been derived for extremal Reissner-Nordstr\"om (RN) black hole using the $AdS_2\times S^2$ geometry of the horizon, a full analysis in the whole geometry is lacking, begging the question of whether it is sufficient to ignore contributions away from the horizon. In this paper, we address this problem and obtain the spatial profile of the Schwinger production rate in an asymptotically flat RN black hole spacetime. We find that the Schwinger effect is strongest on the horizon and decays with distance from the horizon, exhibiting a characteristic scale of the Compton wavelength of the particle. The rate is switched off when the particle's charge-to-mass ratio approaches the corresponding extremality bound for black holes, in accordance with a strong form of the Weak Gravity Conjecture (WGC).}

\begin{document}
\maketitle
\flushbottom

\section{Introduction}

The discussion of effective action in Quantum Electrodynamics dates back to the work of Euler-Heisenberg \cite{Heisenberg:1936nmg} and Weisskopf \cite{Weisskopf:1936hya}, in which polarization of the vacuum due to creation of virtual charged particles from the electromagnetic field is computed. Later on, Schwinger derived the vacuum-persistent amplitude in a constant electric field using the proper time formulation \cite{PhysRev.82.664} and interpreted the imaginary part of the effective action as the production rate of real charged particles\footnote{See \cite{Cohen:2008wz} also regarding this interpretation.}, which is now referred to as the Schwinger effect. These seminal works sparked an interest in experimental proposals to observe the effect (see \cite{Fedotov:2022ely} for a review)  and moreover, opened up a field of theoretical studies of non-perturbative nucleation effects in Quantum Field Theory (QFT).

The Schwinger effect in constant curvature spaces has been derived with various methods in different contexts, for instance with the worldline approach \cite{Affleck:1981bma}, in hyperbolic space \cite{10.1063/1.526781}, AdS space \cite{Pioline:2005pf,Kim:2008xv} and dS space \cite{PhysRevD.49.6343, Frob:2014zka, Kobayashi:2014zza, Hayashinaka:2016qqn}. It has been discussed in (near-extremal) Reissner-Nordstr\"om (RN) black hole spacetime with different asymptotic structures \cite{Gibbons:1975kk,Chen:2012zn,Chen:2016caa,Aalsma:2018qwy, Aalsma:2023mkz}. Despite the existing work, a full account of the spatial dependence of the Schwinger effect has not been presented and therefore the question of the scale of the region relevant for the decay of RN black holes remains to be understood. In this work, we use the worldline instanton approach to compute the spatial profile of the Schwinger production rate of an extremal RN black hole and identify this scale to be the Compton wavelength of the charged particle.

The occurrence of the Compton scale is suggestive of the relation between the Schwinger effect and black hole superradiance, which is originally proposed as a mechanism to extract energy from a Kerr black hole through scattering with an incoming wave. Black hole superradiance involving bosonic fields\footnote{Superradiant amplification of incoming waves by black holes was shown to be absent for fermionic particles, see \cite{Unruh:1973bda,Gueven:1977dq} for instance. This is often considered as a consequence of the Pauli exclusion. For what is relevant to this paper, we note that the existence of stimulated emission of fermions is in no contradiction with absence of superradiance - the latter simply implies that the absorbtion rate is higher than the stimulated emission rate.} was studied in different settings, see \cite{Brito:2015oca} and references therein for a review of the development of the field. The discussion has been extended to the superradiant effect of charged black holes, where extraction of charge and energy of the black hole can occur when the superradiance condition $\omega<q\Phi_H$ is met, where $\omega,q$ are the energy and charge of incoming particle and $\Phi_H$ is the electric potential at the black hole horizon. The superradiance effect have mostly been studied in a first quantized context, which is insufficient to directly capture spontaneous processes like the Schwinger effect. However, the fact that superradiance has a typical extent of, and a rate that is controlled by the Compton wavelength suggests a connection between the Schwinger effect and charged superradiance. They are two sides of a coin - the former is a spontatneous charged radiation process and the latter is a stimulated scattering associated with RN black holes. A connection as such also follows from the long-standing principle of detailed balance, from which a first establishment between spontaneous and stimulated emission of a simple quantum system was obtained by Einstein. Soon after the realization that black holes emit Hawking radiation \cite{Hawking:1974rv, Hawking:1975vcx, Wald:1975kc, Unruh:1976db}, an early application of this principle to black holes was presented by Wald \cite{Wald:1976ka}, where it was shown that the stimulated emission of neutral particles implies spontaneous emission. The same logic can be generalized to charged black holes, which indicates that stimulated charged emission of black holes is always associated with a spontaneous emission process. We identify the former as charged superradiance and the latter as the Schwinger effect.

The study on decay of charged black holes has deep relations to the Weak Gravity Conjecture (WGC) originally proposed in \cite{Arkani-Hamed:2006emk}. (See \cite{Harlow:2022ich, Montero:2024qml} for a general review.) In its simplest form, the conjecture requires the existence of at least one superextremal particle with charge-to-ratio larger than the corresponding black hole extremality bound. The superextremal particles can lead to the decay of non-supersymmetric extremal black holes, whose discharge is constrained to prevent exposure of naked singularities. The WGC can take different forms in spacetimes with different asymptotic structures. One interesting direction is to obtain the form of the WGC bound in dS space. Since the decay of extremal black holes is linked to the WGC, understanding the Schwinger production can be of great benefits to identifying the WGC bound in different settings. For instance, the Schwinger rate we computed for an extremal black hole in asymptotically flat space registers the information of the extremality bound - the rate sees a switch-off behavior when the charged particle tends to extremality from above. This is in accordance with a strong form of WGC in flat space and leads to the speculation that the Schwinger rate might be indicative of the WGC bound for general cases. While in this paper we do not present concrete results on the Schwinger effect of RN black holes in dS space, we note the possibility of generalizing the worldline approach adopted in this paper to the study of dS black holes. We leave this interesting question to be tackled in a future work.

The paper is organized as follows: In section \ref{worldline_formalism}, we review the worldline path integral formalism used to compute the Schwinger pair production rate. We devote section \ref{BH_Schwinger_rate} to the computation of the instanton paths, instanton action and one-loop determinant in the extremal RN black hole spacetime, obtaining the local Schwinger production rate in the exterior of the black hole. In section \ref{conclusions}, we summarize our findings and conclude that the radial profile of the Schwinger effect is characterized by the Compton wavelength of the particle and that the production is switched off when the particle's charge-to-mass ratio tends to the extremality bound for charged black holes in flat space. We further discuss the connection between the Schwinger effect and black hole superradiance and the implications of Schwinger effect to bounds on the particle spectrum.

\section{Effective action and worldline instantons} \label{worldline_formalism}
In this section, we review the formalism for computing the Schwinger effect. The creation rate for charged particles in an electric field was first derived in \cite{PhysRev.82.664}, where the production rate $\Gamma$ is expressed in terms of a regularized vacuum amplitude,
\beq
\Gamma=1-\frac{|\braket{0_A|0_A}|^2}{|\braket{0|0}|^2}.
\eeq
The quotient $\frac{|\braket{0_A|0_A}|^2}{|\braket{0|0}|^2}$ concerns two vacuum states $\ket{0_A}, \ket{0}$ defined with and without the gauge field\footnote{The choice of time-coordinate in curved space implicitly defines these vacua, which are the surviving projection in the past and future infinity.} and it has the meaning of vacuum persistence. In other words, $\Gamma$ is the probability that the system is not in the vacuum state due to the gauge field. The vacuum-to-vacuum amplitudes are written as path integrals
\beq \label{general_action}
\frac{\braket{0_A|0_A}}{\braket{0|0}}=\frac{\int \mc{D}\{\psi\} e^{iS[\{\psi\},A]}}{\int \mc{D}\{\psi\} e^{iS[\{\psi\}]}},
\eeq
where $\{\psi\}$ is used to generally represent all the fields that are integrated out and $S$ is the action associated with these fields. Eq.(\ref{general_action}) defines the effective action $e^{iS_\text{eff}[A]}\equiv \frac{\braket{0_A|0_A}}{\braket{0|0}}$ whose imaginary part is related to the Schwinger production rate by
\beq
\Gamma=2\text{Im}(S_\text{eff}).
\eeq

The specific field content that we consider in this paper is a complex scalar field charged under a U(1), denoted as $\phi(x)$. The scalar field action is assumed to be quadratic in $\phi$, given by
\beq \label{quadratic_action}
S=\int \sqrt{-g} d^4x\phi^* H_A \phi,
\eeq
where $\sqrt{-g}d^4x$ is the curved space volume form, $H_A=D_\mu D^\mu -m^2$ is a Hermitian operator and $D_\mu=\nabla_\mu+ieA_\mu$ is the covariant derivative in curved space containing also the gauge connection. The gravitational field and gauge field are taken to be fixed, so the associated Einstein-Hilbert and Maxwell action terms contribute only a constant phase to the vacuum amplitude that does not affect the production rate. For this reason, the two action terms will be omitted in discussion of the Schwinger effect. Further, the form of Eq.(\ref{quadratic_action}) and the gauge field being non-dynamical means that the effective action is generated by all the one-loop diagrams with arbitrary external gauge field legs and scalar field propagating in the loop.

The path integral over the complex scalar field is Gaussian due to the quandratic nature of its action, which can be formally expressed as a functional determinant
\beq
\int \mc{D}\phi^*\mc{D}\phi e^{-S_E}=(\det{H_A})^{-1},
\eeq
where we have chosen to proceed in Euclidean time. Introducing the auxiliary proper time parameter and applying the log-determinant identity, the abstract determinant is put into the form of a proper time integral of the kernel $\braket{x|e^{sH_A}|x}$
\beq \label{trace}
\ln \det{H_A}=\text{tr}\ln H_A=\text{tr}\int \frac{ds}{s}e^{sH_A}=\int \sqrt{-g} d^4x \int \frac{ds}{s}\braket{x|e^{sH_A}|x}.
\eeq

A common approach to evaluate the above integral is to perform the trace in the eigenvalue basis of the operator $H_A$, which involves a sum over the eigenstates 
\beq \label{heat_kernel}
\int d^4x\int_0^\infty \frac{ds}{s}\int d\lambda |\phi_\lambda(x)|^2 e^{-s\lambda}.
\eeq
Here $\lambda$ and $\phi_\lambda$ are the eigenvalue and eigenfunction of the operator $H_A$ respectively. This is referred to as the heat kernel method \cite{DeWitt:1964mxt,PhysRev.162.1239,Vassilevich:2003xt} which has been applied to computations of effective action in hyperbolic space \cite{Pioline:2005pf, 10.1063/1.526781} and black hole entropy corrections \cite{Sen:2007qy,Banerjee:2010qc,Sen:2012dw,Sen:2012cj,Shiu:2016weq}.

There are manifest divergences in Eq.(\ref{heat_kernel}) with different physics origins. The UV divergence at small $s$ is related to the renormalization of the couplings in the theory. The operator $H_A$ can admit zero modes with $\lambda=0$, leading to an IR divergence at large $s$. The zero modes present in the derivation of the Schwinger rate in flat space and $AdS_2$ arise from the spacetime symmetry - all points in constant curvature space being equivalent. They appear to be pure gauge and the treatment is to extract and replace them by collective coordinates, resulting in a volume factor upon integration. The Schwinger effect of particle creation is closely related to a different type of IR divergence in the heat kernel expression - negative modes. The coupling of the scalar field to the background gauge field shifts the spectrum of the operator $H_A$ and generates negative modes that would have been absent. The negative mode indicates an instability of the system under influence of the background field. For the Schwinger effect, the electric field triggers the transition of the original vacuum state to a state with non-zero particle occupation number. Unlike the pure gauge modes, negative modes contain important physical information of the decay process and thus require special care when being dealt with. It is a goal of this paper to accomplish this in the setting of extremal RN black holes using the worldline path integral formalism, which we review in the following paragraphs.

A central idea of the worldline formalism is to rewrite the kernel $\braket{x|e^{-sH_A}|y}$ as a quantum mechanical position space integral over all paths $\xi(\tau)$ connecting $x$ and $y$. The formalism was inspired by one-loop vacuum amplitude computations in string theory (see \cite{Reuter:1996zm} for an overview) and was developed into an alternative method \cite{Strassler:1992zr} to Feynman diagrams for computations of effective actions in QFT. It had seen use in the derivation of the Schwinger effect in flat space \cite{Affleck:1981bma, PhysRevD.49.6327,  Dunne:2005sx} and was discussed in the context of curved space and higher spins \cite{Bastianelli:2002fv, Bastianelli:2005rc,Bastianelli:2006rx,Corradini:2015tik}. Of course, the worldline method is based on an earlier idea of path integrals \cite{Feynman:1948ur}, whose generalization to curved space can be found in \cite{DeWitt:1957at}.
While the perturbative expansion of the worldline action in curved space and its renomalization have been studied \cite{Bastianelli:2002fv}, an application of the formalism to a black hole spacetime has not been put forth. We achieve the goal of computing the non-perturbative\footnote{Here, we mean a result that is non-perturbative in the coupling constant $e$.} Schwinger rate in the extremal RN spacetime by applying the stationary point approximation to the worldline path integral.

Using the worldline formalism, we expressed the kernel in Eq.(\ref{trace}) as\footnote{This form holds for spacetimes with vanishing Ricci scalar.}
\beq \label{propagator}
\braket{x|e^{H_As}|x}=\int_{\xi(0)=x}^{\xi(s)=x}\mc{D}\xi e^{-\int_0^s d\tau [\frac{1}{4}(\frac{D\xi}{d\tau})^2+e\int A_\mu d\xi^\mu+sm^2]}.
\eeq
$S_{wl}\equiv \int_0^s d\tau [\frac{1}{4}(\frac{D\xi}{d\tau})^2+e\int A_\mu d\xi^\mu+sm^2]$ is interpreted as the worldline action\footnote{In some papers, the gauge field related action term has an $i$ in front, but the Euclidean gauge field is imaginary. We choose to absorb the imaginary unit into the gauge field. The different conventions lead to the same equations of motion.} and $\xi$ as the worldline parameterized by its proper time $\tau$. The auxiliary parameter $s$ now has the meaning of the total proper time of the worldline. It appears to be more convenient to uniformly parametrize the total proper time of the worldlines as 1, motivating the following rescaling $s\rightarrow\frac{s}{m^2}$ and $\tau\rightarrow \frac{s}{m^2}\tau$. Eq.(\ref{trace}) becomes
\beq \label{rescaled_action}
\int \sqrt{g} d^4x \int \frac{ds}{s} \int_{\xi(0)=x}^{\xi(1)=x} \mc{D}\xi e^{-[s+\frac{m^2}{4s}\int_0^1d\tau (\frac{D\xi}{d\tau})^2+e\int d\tau A_\mu \frac{D\xi^\mu}{d\tau}]}.
\eeq
The advantage of rewriting the kernel in this form is that the UV divergence in the small $s$ regime corresponding to high momentum paths is manifestly regularized by the kinetic term of the worldline action as seen from Eq.(\ref{rescaled_action}).  

Proceeding from here, one has different choices in what order the integrals in $s$ and $\xi$ are performed. In \cite{Affleck:1981bma}, the proper time integral was first done using a stationary point approximation, which generates a non-local term for the paths $\xi$ in the worldline action. The non-local term hinders the further evaluation of the one-loop determinant of the path fluctuations. Another route, taken in \cite{Dunne:2006st}, was to first consider the integration over the paths. This will generate an $s$-dependent term which would then be combined with the remainder in Eq.(\ref{rescaled_action}) to be integrated. This choice is less applicable when the integration over the paths does not yield a fully analytical expression. We will instead perform the stationary point approximation to both the $s$ and $\xi$ integrals simultaneously, identifying the stationary points in the space of $(s,\xi)$. The final result should be independent of this choice because the order of evaluation only alters the basis in $(s,\xi)$, not affecting the special points of the worldline action and the determinant of its second variation.

The worldline action is expanded around the stationary points $(\Bar{s},\Bar{\xi})$ up to second order, see Appendix \ref{action_expansion}. The stationary points are determined by imposing vanishing first variations,
\beq \label{EoM} \bc
\Bar{s}=\frac{m}{2}\sqrt{\int_0^1d\tau g_{\mu\nu}\Dot{\Bar{\xi}}^\mu\Dot{\Bar{\xi}}^\nu}\\
\left(\frac{m^2}{2s}g_{\mu\nu} \frac{D^2}{d\tau^2}-eF_{\mu\nu}\frac{D}{d\tau}\right)\Bar{\xi}^\nu=0
\ec. \eeq
To simplify notation, the overbar will be omitted and the restriction to the stationary points will be assumed unless otherwise stated. The second order expansion in Appendix \ref{action_expansion} can be expressed compactly as
\beq \label{hessian}
S^{(2)}=\frac{1}{2}\left(\delta s \mf{H}_{ss} \delta s +\delta s \mf{H}_{s\xi} \delta\xi +\delta\xi \mf{H}_{\xi s} \delta s +\delta\xi \mf{H}_{\xi\xi}\delta\xi\right),
\eeq
where $\mf{H}$ is the Hessian operator of the worldline action. Note that an integration over $\tau$ is implicit in the definition of how $\mf{H}$ acts on the path fluctuations, which is explicitly given in Appendix \ref{action_expansion}. To facilitate the computation of the one-loop determinant of the second variation, we diagonalize the Hessian with respect to the path fluctuations
\beq
S^{(2)}=\frac{1}{2}\left( \delta s \tilde{\mf{H}}_{ss}\delta s+\delta\xi'\mf{H}_{\xi\xi}\delta\xi'\right)
\eeq
where the diagonalized element, denoted by $\tilde{\mf{H}}_{ss}$, and the shifted path fluctuation is defined as
\beq \label{diagonalize} \bc
\tilde{\mf{H}}_{ss}=\mf{H}_{ss}-\mf{H}_{s\xi}\mf{H}_{\xi\xi}^{-1}\mf{H}_{\xi s}\\
\delta\xi'=\delta\xi +\mf{H}_{\xi\xi}^{-1}\mf{H}_{\xi s}\delta s
\ec. \eeq
More explicitly, $\tilde{\mf{H}}_{ss}$ is given by
\beq \label{diagonalized_Hess}
\int_0^1d\tau\frac{m^2}{4s}g_{\mu\nu}\Dot{\xi}^\mu\Dot{\xi}^\nu-\int_0^1 d\tau \int_0^1d\tau'\frac{m^2}{2s^2}g_{\mu\alpha}\frac{D^2}{d\tau^2}\xi^\alpha(\tau)G^{\mu\nu}(\tau,\tau')\frac{m^2}{2s^2}g_{\nu\beta}\frac{D^2}{d{\tau'}^2}\xi^\beta(\tau').
\eeq
In Eq.(\ref{diagonalize}), an operator inverse is involved. It is the matrix-valued Green's function associated with the operator $\mf{H}_{\xi\xi}$. To simplify the notation in section \ref{BH_Schwinger_rate}, we denote $\Lambda\equiv \mf{H}_{\xi\xi}$ and $G\equiv\mf{H}_{\xi\xi}^{-1}$. The Green's function satisfies the following differential equation
\beq
\Lambda G =I \delta(\tau-\tau'),
\eeq
with $I$ being the indentity matrix with the same rank as $\Lambda$. The Green's function is defined together with the boundary conditions. From Eq.(\ref{rescaled_action}), the requirement that the paths begin and end at the same point suggests the boundary conditions on the path fluctuations be Dirichlet, $\delta\xi(0)=\delta\xi(1)$. Therefore, the Green's function by definition satisfies the same boundary conditions in both variables $\tau, \tau'$. This ensures that the shifted path fluctuation $\delta\xi'$ also satisfies the Dirichlet condition, $=\delta\xi'(0)=\delta\xi'(1)=0$. We give the construction of $G$ in Appendix \ref{Greens_function}.

In the next few paragraphs, we analyze the stationary conditions for the stationary points. We observe that Eq.(\ref{EoM}) is the geodesic equation of a charged particle coupled to an electric field in curved space with a rescaled mass parameter $m^2\rightarrow \frac{m^2}{4s}$, if we interpret $\dot{\xi}^\mu\equiv\frac{D}{d\tau}\xi^\mu$ as the particle velocity along the worldline. For this reason, following \cite{Affleck:1981bma}, we will call these solutions worldline instantons, or instanton paths. The geodesic equation is a second order differential equation, from which a set of two first order equations, Eq.(\ref{velocity}) and Eq.(\ref{first_integral}), can be obtained as first integrals. Contracting the second equation in (\ref{EoM}) with the particle velocity and making use of the anti-symmetric property of $F_{\mu\nu}$, one obtains $\frac{D}{d\tau}(\Dot{\xi})^2=0$, so
\beq \label{velocity}
g_{\mu\nu}\Dot{\xi^\mu}\Dot{\xi^\nu}=a^2=\text{const},
\eeq
and $\Bar{s}=\frac{ma}{2}$ from Eq.(\ref{EoM}).
Choosing a static gauge $A=A_0(r) dt$, the geodesic equations can be written as
\beq \bc \label{geodesic_eqn}
\frac{m}{a}\frac{D^2t}{d\tau^2}=eF^0_{\;\;1}\frac{Dr}{d\tau}\\
\frac{m}{a}\frac{D^2r}{d\tau^2}=-eF^1_{\;\;0}\frac{Dt}{d\tau}
\ec, \eeq
We should note that several assumptions were made when obtaining the above equations: we are considering a four dimensional spherically symmetric spacetime with coordinates $\xi^\mu=(t,r,\theta,\varphi)$. We also imposed the condition that the particles have no angular motion. The latter assumption is reasonable since we are identifying the stationary points that dominate the worldline path integral and an addition of angular momentum will increase the worldline action, suppressing the contribution to the path integral. The static gauge choice allows for a first integral of the first line in Eq.({\ref{geodesic_eqn}), which yields a conserved quantity $\omega$ that can be understood as the energy of the particle
\beq \label{first_integral}
\omega=\tilde{m}g_{00}\Dot{t}+eA_0.
\eeq
Combining the above expression with Eq.(\ref{velocity}), the radial coordinate separates from the time coordinate and we obtain
\beq \label{radial_ode}
\Dot{r}=\pm a\sqrt{g^{-1}_{11}\left[1-\frac{(eA_0-\omega)^2}{m^2g_{00}}\right]}.
\eeq
Eq.(\ref{first_integral}) and Eq.(\ref{radial_ode}) will be the key equations used to compute the instanton paths.

Finally, we define the local effective action as the integrand of Eq.(\ref{rescaled_action}) inside the spacetime integral,
\beq \label{local_action} \bal
w(x)&=\int_{\xi(0)=x}^{\xi(1)=x}\mc{D}\xi\int_0^\infty\frac{ds}{s} e^{-S[s,\xi]}\\
&= \mc{A}(\Bar{\xi}) e^{-S^{(0)}[\Bar{s},\Bar{\xi}]}
\eal \eeq
where $\Bar{\xi}$ is the stationary path starting and ending at $\xi=x$, $S^{(0)}$ is the corresponding stationary worldline action and $\mc{A}(\Bar{\xi})=\Bar{s}^{-1}(\det{\mf{H}})^{-\frac{1}{2}}=\Bar{s}^{-1}(\tilde{\mf{H}}_{ss}\det{\Lambda})^{-\frac{1}{2}}$. The number of negative modes of the operator $\Lambda$ determines the phase of the local effective action. The local Schwinger rate corresponds to the imaginary part of the effective action, and with exactly one negative mode, it can be expressed as
\beq
\Gamma(x)=\text{Im}[w(x)]=|\mc{A}(\Bar{\xi})| e^{-S^{(0)}[\Bar{s},\Bar{\xi}]}.
\eeq
For the case of Schwinger effect in constant curvature space, factoring out the spacetime integral and defining the volume rate is a necessary step towards a physical and finite answer of the particle production rate. A local definition of the production rate extends beyond the homogeneous case since the electric field induced particle creation is clearly a local effect - an observer can place a detector at a specific location in the electric field and expect charged particles to be detected at some rate. We would like to understand the spatial profile of the particle production rate in the extremal RN black hole spacetime.

\section{Schwinger production rate of charged black holes} \label{BH_Schwinger_rate}
In this section, we compute the Schwinger production rate using the worldline formalism reviewed in section \ref{worldline_formalism}. The stationary points are first obtained, from which we compute the stationary worldline action and the associated prefactor. The latter is achieved by finding the determinant of the second variation operator evaluated at the stationary points.

Before we dive into the computation of the Schwinger rate, it is beneficial to first identify the physically relevant scenarios for the production process. By that, we refer to the particle spectrum that enables 
the decay of extremal black holes. While the analysis in this paper is done in a fixed background without back-reactions on the geometry, the consequence of the back-reactions should not be overlooked. In the allowed parameter space $(Q,M)$ for black holes with charge $Q$ and mass $M$ in flat space, extremal RN black holes lie at the boundary of the valid space $Q\le M\ell_P$ where no naked singularity exists.  In asymptotically flat space, the charge and energy of the gravitating system is well defined at the asymptotic boundary and the charge and mass of a produced particle is subtracted from the black hole when it reaches the asymptotic region. The black hole extremality bound therefore only permits particles with $q>m$ to escape to infinity and discharge the black hole. This can be shown more concretely by considering the reversed process of the thought experiment in \cite{Wald:1974hkz}. However, a classical picture suffices to demonstrate this point. Away from an RN black hole, the gravitational and electric potential is well described by the inverse power law, generating for a particle with charge $q$ and mass $m$ a total potential of the form
\beq
V(r)\sim \frac{qQ-mM\ell_P^2}{r}.
\eeq
For superextremal particles with $\frac{e}{m\ell_P}>\frac{Q}{M \ell_P}=1$, the potential is repulsive and the particle will be accelerated by the electric field after nucleation with near zero kinetic energy around the black hole and reach infinity, reducing the black hole to a slightly subextremal one. On the contrary, if the particle has $\frac{e}{m\ell_P}<1$, the potential is attractive and the particle will eventually be re-absorbed by the black hole. In the latter situation, an asymptotic observer will not see a flux of particles coming from the black hole. While this does not rule out the possibility that subextremal particles cannot locally or temporarily exist around the black hole, for the purpose of computing the Schwinger effect, we will only be interested in the production rate for particles with $z=\frac{e}{m\ell_P}>1$ that lead to the decay of the black hole.

\subsection{Instanton solutions in Extremal RN black hole spacetime}
\label{stationary_path}

We consider an extremal RN black hole background with metric
\beq
ds^2=fdt^2+\frac{dr^2}{f}+r^2(d\theta^2+\sin^2\varphi^2),
\eeq
and gauge field
\beq
A=\frac{Q}{r}dt,
\eeq
where $f=(1-\frac{Q\ell_P}{r})^2$. Throughout this paper, we use the Planck length $\ell_P=\frac{1}{\sqrt{G_N}}$ as the only dimensionful unit\footnote{Further, the $4\pi$ factor and vacuum permittivity are absorbed into the definition of the charge and coupling constant. The complex scalar particle we consider is assumed to carry one charge $q=1$.}.

The radial trajectory is governed by Eq.(\ref{radial_ode}) particularized on the RN spacetime
\beq \label{radial_eqn_0}
\Dot{r}=\pm a \sqrt{\left(1-\frac{Q\ell_P}{r}\right)^2-\frac{e^2}{m^2}\left(\frac{Q}{r}-\frac{\omega}{e}\right)^2}.
\eeq
To put the above equation into a more convenient form, we make the change of the radial coordinate $\rho=\frac{Q\ell_P}{r}$, denote the charge-to-mass ratio of the particle as $z=\frac{e}{m\ell_P}$ and define the rescaled energy parameter  $\rho_0=\frac{\omega \ell_P}{e}$. The new radial coordinate maps the black hole exterior to $\rho\in(0,1]$. We further denote the function under the square root of Eq.(\ref{radial_eqn_0}) as $h(\rho)=(1-\rho)^2-z^2(\rho-\rho_0)^2$, which corresponds to an effective potential in the radial direction.  We can now rewrite Eq.(\ref{radial_eqn_0}) as
\beq \label{radial_eqn}
\Dot{\rho}=\mp \frac{a}{Q\ell_P}\rho^2\sqrt{h(\rho)}.
\eeq

The particles that we will be concerned with, as explained at the beginning of this section, are superextremal with $z>1$. In Euclidean signature, the coordinates are space-like, and consequently the gauge field is magnetic-like. The trajectories of the charged particles are spirals, whose instantaneous radii depend on the local field strength. The paths that contribute to the effective action has the same beginning and ending points according to Eq.(\ref{rescaled_action}), which translates to the condition that the radial and time components of the paths admit turn-around points. This gives rise to a constraint on the instanton parameter $\rho_0\in (0,1)$. The full trajectory is schematically shown in Fig.\ref{spiral} and the contributing fraction of the path is the self-intersected loop.

\begin{figure}[htbp]
    \centering
    \includegraphics[width=10cm]{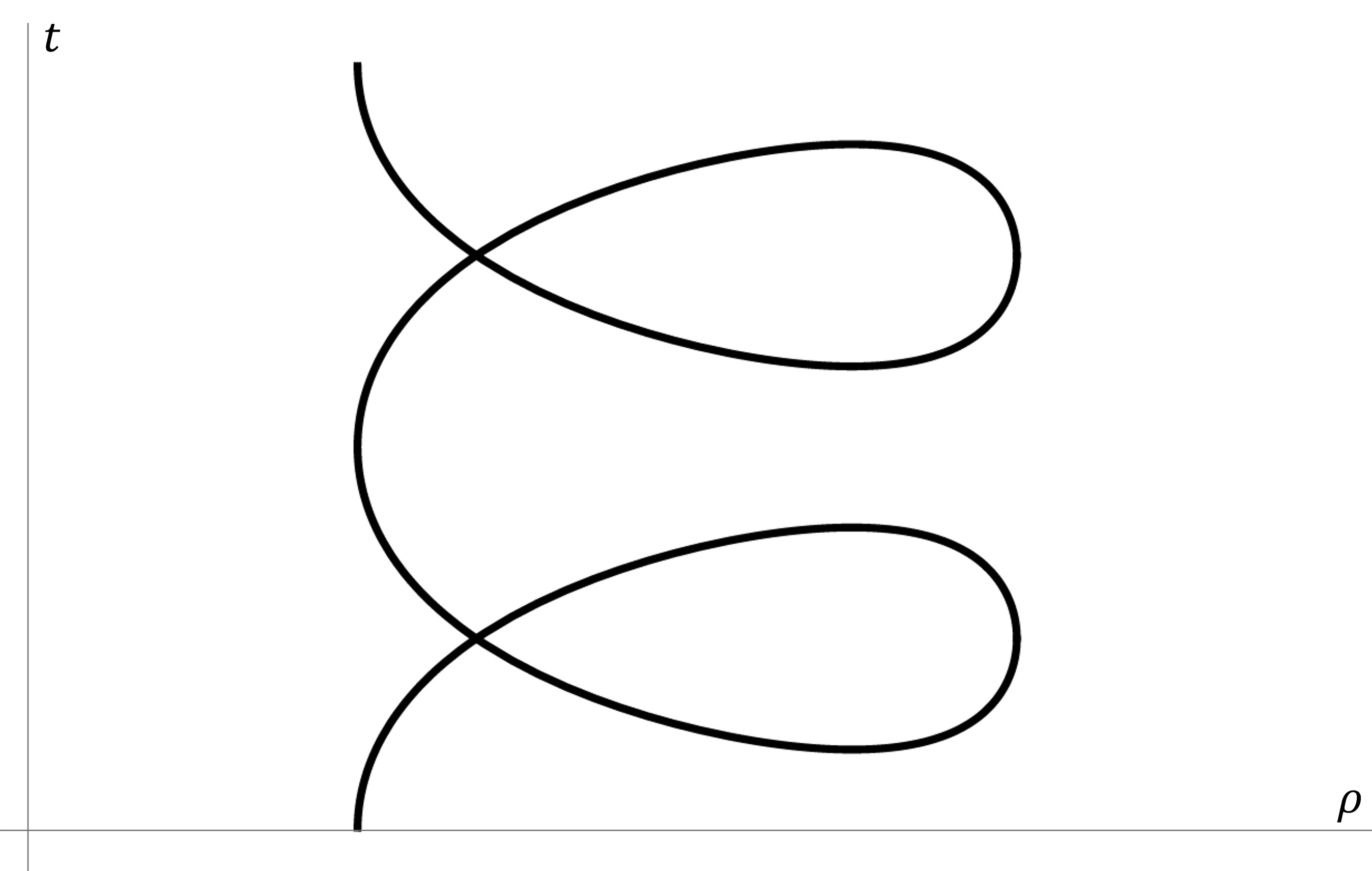}
    \caption{The instanton path. The full path is a spiral that extends infinitely in the $t$ direction and goes back and forth between the two turn-around points $\rho_{1,2}$ in the $\rho$ direction. The path self-intersects at $\rho=\rho_\times$ and the loop contributes to the effective action through the path integral.}
    \label{spiral}
\end{figure}

By direct integration, the solution to Eq.(\ref{radial_eqn}) is found to be 
\beq \label{ext_radial_sol}
\frac{\sqrt{(\rho_1-\rho)(\rho-\rho_2)}}{\rho_1\rho_2\rho}+\frac{\rho_1+\rho_2}{(\rho_1\rho_2)^\frac{3}{2}}\arctan{\sqrt{\frac{\rho_1(\rho_2-\rho)}{\rho_2(\rho-\rho_1)}}}=\frac{a\sqrt{z^2-1}}{Q\ell_P}(\tau-\tau_0),
\eeq
where $\tau_0$ is an integration constant and $\rho_{1,2}=\frac{z \rho_0\pm 1}{z\pm 1}\in (0,1]$ are the two zeros of $h(\rho)$. Eq.(\ref{ext_radial_sol}) implicitly defines the function $\rho(\tau)$. To identify the intersection point, we solve for $t(\tau)$, which is governed by Eq.(\ref{first_integral}). In the redefined coordinates and parameters, the equation appears as
\beq
\Dot{t}=\frac{za(\rho_0-\rho)}{f(\rho)}.
\eeq
It is convenient to switch from $\tau$ to the variable $\rho$ using the identity $\Dot{t}=\Dot{\rho}\frac{dt}{d\rho}$ together with Eq.({\ref{radial_eqn}), leading to
\beq \label{dtdrho}
\frac{dt}{d\rho}=\mp Q\ell_P z\frac{\rho_0-\rho}{\rho^2f\sqrt{h(\rho)}}.
\eeq

Denoting the endpoints of the path by $\rho(0)=\rho(1)=\rho_\times, t(0)=t(1)=0$, we determine $a$ and $\rho_\times$, which specifies the instanton paths, by solving the following equations:
\beq \bc
\rho(\tau=0)=\rho(\tau=1)=\rho_\times\\
\rho(\tau=\frac{1}{2})=\rho_2\\
t(\tau=0)=t(\tau=\frac{1}{2})=0
\ec. \eeq
Because the intersection $\rho_\times$ is a root of a transcendental equation, in general the paths will have to be computed numerically.

\subsection{The instanton action}
We next compute the worldline instanton action. This is given by the sum of a kinetic term $ma$ and a gauge field associated term
\beq \bal
S_A&=e\int_0^1 d\tau\frac{A_0(\omega-eA_0)}{\tilde{m}g_{00}}=\frac{e^2a}{m\ell_P^2}\oint_\gamma \frac{d\rho}{\Dot{\rho}}\frac{\rho(\rho_0-\rho)}{(1-\rho)^2}\\
&=-Qm\ell_P\frac{z^2}{\sqrt{z^2-1}}\times 2I,
\eal \eeq
where
\beq \label{complex_intg}
I=\int_{\rho_\times}^{\rho_2}F(\rho)d\rho\equiv\int_{\rho_\times}^{\rho_2}\frac{(\rho_0-\rho)d\rho}{\rho(1-\rho)^2\sqrt{(\rho_2-\rho)(\rho-\rho_1)}}.
\eeq
The same technique used to obtain Eq.(\ref{dtdrho}) is applied here to switch the integration variable to $\rho$. Generally outside the black hole, the lower limit $\rho_\times$ is the solution to a transcendental equation and has to be determined numerically. Making use of the numerical solutions obtained in section \ref{stationary_path}, we compute the instanton action and obtain the full profile of the exponential term of the Schwinger rate outside the extremal RN black hole. The result is presented in Fig.\ref{action}.

\begin{figure}[htbp]
    \centering
    \includegraphics[width=14cm]{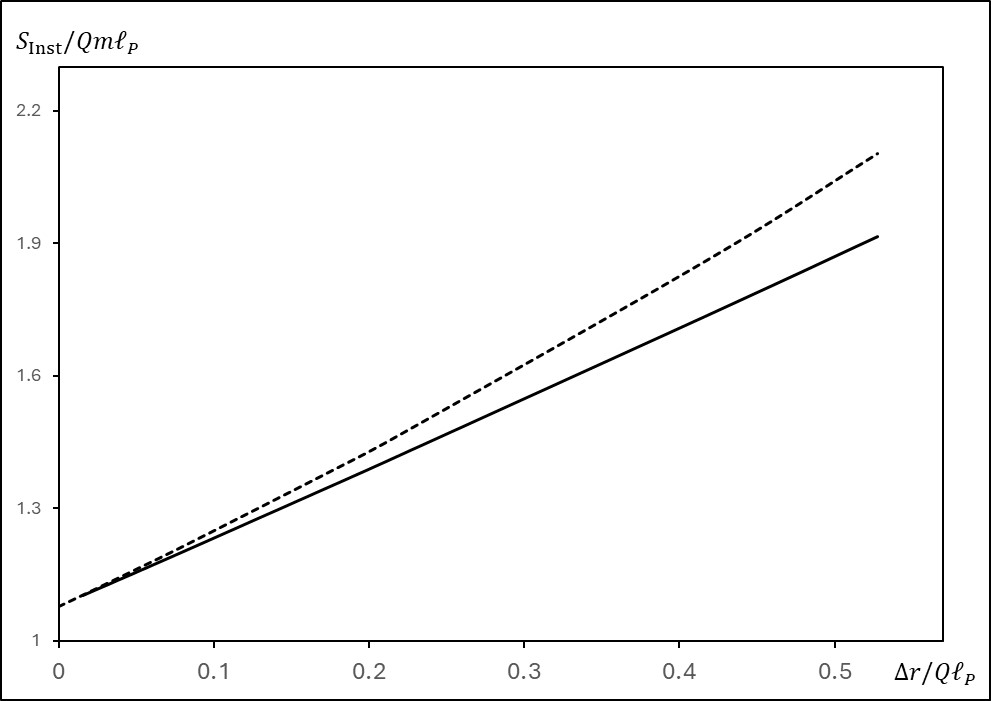}
    \caption{The radial dependence of the instanton action. $\Delta r$ is the distance to the horizon. The solid line is computed numerically and the dotted line is computed according to Eq.(\ref{ext_instanton_action}). The numerical result is well described by the analytical approximation close to the horizon.}
    \label{action}
\end{figure}

To check consistency with existing results in $AdS_2$, we seek an approximation of the action integral near the horizon. We first note that $\rho_0\rightarrow 1$ translates to the near horizon limit since the two turn-around points $\rho_{1,2}=\frac{z \rho_0\pm 1}{z\pm 1}\rightarrow 1$ both tend to the horizon in this limit. We then observe that $\frac{\rho_\times - \rho_1}{\rho_2-\rho_1}\rightarrow 0$ as the instanton path approaches the horizon. This means that in the near horizon region, Eq.(\ref{complex_intg}) is well approximated by the same integral but with the lower limit $\rho_\times$ replaced by $\rho_1$. In fact, as the horizon is approached, the instanton paths, after proper coordinate transformation, tend to those obtained in $AdS_2$, which are closed trajectories discussed in Appendix \ref{AdS-instanton}. Setting $\rho_\times\rightarrow \rho_1$, Eq.(\ref{complex_intg}) can then be analytically integrated using the residue theorem. The contour is chosen to wrap around infinity and the branch cut between $\rho=\rho_{1,2}$, shown in Fig.\ref{contour}. The residues of $F(\rho)$ at poles $\rho=0,1$ are
\beq \bc
\text{Res}[F(\rho=0)]=\frac{i\rho_0}{\rho_1\rho_2}\\
\text{Res}[F(\rho=1)]=\frac{i\rho_0}{2}\frac{(2\rho_1\rho_2-3\rho_1-3\rho_2+4)}{[(1-\rho_1)(1-\rho_2)]^\frac{3}{2}}-\frac{i}{2}\frac{(2-\rho_1-\rho_2)}{[(1-\rho_1)(1-\rho_2)]^\frac{3}{2}}
\ec. \eeq
Only the contour integrals along $\mc{C}_{1,3}$ have a non-zero contribution and each turns out to contribute $I$, therefore
\beq \label{I}
2I=2\pi \left\{\frac{\rho_0}{\sqrt{\rho_1\rho_2}}+\frac{\rho_0(2\rho_1\rho_2-3\rho_1-3\rho_2+4)-(2-\rho_1-\rho_2)}{2[(1-\rho_1)(1-\rho_2)]^\frac{3}{2}}\right\}.
\eeq
Substituting the turn-around points with the instanton parameter $\rho_{1,2}=\frac{z \rho_0\pm 1}{z\pm 1}$, we obtain
\beq
2I=2\pi\frac{\sqrt{z^2-1}}{z}\left(-1+\frac{z \rho_0}{\sqrt{z^2\rho_0^2-1}}\right),
\eeq
and the total action reads
\beq \label{ext_instanton_action}
S_\text{inst}=ma+S_A=2\pi Qm\ell_P\left[\frac{z^2\rho_0-1}{(z^2\rho_0^2-1)^{\frac{3}{2} }}+z-\frac{z^2\rho_0}{\sqrt{z^2\rho_0^2-1}}\right].
\eeq
From Eq.(\ref{ext_instanton_action}), it is easy to recover the $AdS_2$ instanton action by taking the limit $\rho_0\rightarrow 1$, and one finds that
\beq
\lim_{\rho_0\rightarrow 1} S_\text{inst}=2\pi Qm\ell_P(z-\sqrt{z^2-1}),
\eeq
in agreement with \cite{Pioline:2005pf} and the computations in Appendix \ref{AdS-instanton}.

Using the above approximate form, near the horizon the instanton action can be expressed as a dimensionless ratio of the black hole size and the particle's Compton wavelength $\lambda_c=m^{-1}$,
\beq \label{approx_action}
S\sim \frac{Q\ell_P}{c(z)\lambda_c},
\eeq
where $c(z)$ is a $z$-dependent parameter. When $z$ is large, $c(z)\sim z$ and when $z$ is of order unity, $c(z)\sim O(1)$.

The stationary point approximation applied to obtain this result requires $S\gg 1$. There are two independent parameters that one can dial to explore the boundary where the action is of order unity $S\sim 1$ - this is the point where the production rate $\Gamma\propto e^{-S}$ is no longer suppressed and the stationary point approximation starts to break down. From Eq.(\ref{approx_action}) we can see that the particle production process will tend to be unsuppressed when (a) the particle becomes very light ($m\rightarrow 0$ with $z$ or $e$ kept fixed), or (b) the charge-to-mass ratio is very high ($z\rightarrow \infty$ with $m$ kept fixed).

To put in some context, consider electrons whose $z\sim 10^{22}$ and $\lambda_c\sim 10^{-12}m$, and a solar mass charged black hole of radius $r_S\sim 10^3m$. The action is roughly $S\sim 10^{-7}$, indicating no suppression of the Schwinger process and thus a short lifetime of an extremally charged black hole due to rapid discharge. This shows why we do not expect to see black holes carrying high charges in nature. If we consider a particle charged under a different (hidden) U(1) sector with $z\sim 1$, then the extremal black hole carrying the hidden U(1) charge can potentially have a longer lifetime, if $Qm\ell_P\gg 1$.

The calculation we present concerns an electrically charged black hole, but the analysis and results hold as well if one considers magnetic charge emission of magnetic black holes. For instance, if we consider the 't Hooft-Polyakov magnetic monopole \cite{tHooft:1974kcl,Polyakov:1974ek}, the charge and mass of the monopole are related to the electric coupling and the cut-off scale $\Lambda_c$ of the theory by
\beq
q_\text{mag}\sim \frac{1}{e}, \;\; m_\text{mag}\sim \frac{\Lambda_c}{e^2}.
\eeq
The charge-to-mass ratio of the magnetic monopole is estimated as $z_\text{mag}\sim \frac{e}{\Lambda_c \ell_P}$ and the Compton length will be $\lambda_{c,\text{mag}}\sim \frac{e^2}{\Lambda_c}$. If we take the cut-off scale to be the GUT scale, then for a solar sized black hole, $S\sim 10^{33}$ and the Schwinger effect for magnetic charge production will be highly suppressed. For a lower cut-off scale such as the electroweak scale, magnetic charge production of sub-solar sized black holes with mass $M\sim 10^{-3} M_\odot$ can potentially be cosmologically relevant.

\begin{figure}[htbp]
    \centering
    \includegraphics[width=10cm]{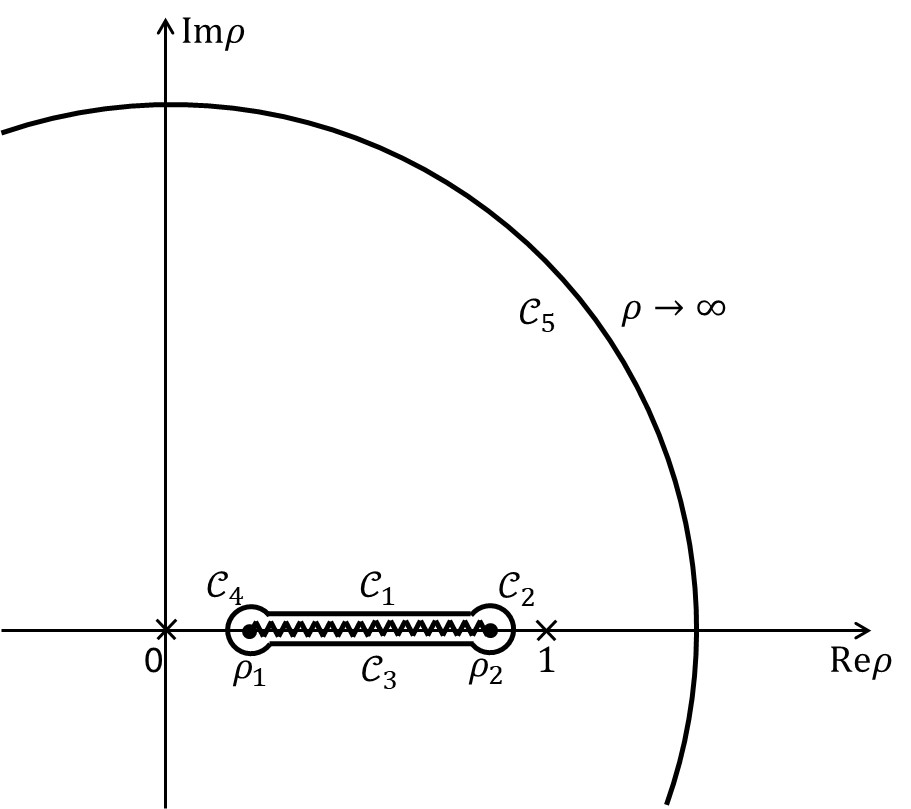}
    \caption{The contour for integration of Eq.(\ref{complex_intg}). $F(\rho)$ has poles at $\rho=0,1$ and branch points at $\rho=\rho_{1,2}$. The branch cut is chosen to connect the branch points along the real axis. The contribution from contour $\mc{C}_{2,4}$ and $\mc{C}_5$ vanishes. The contribution from $\mc{C}_{1,3}$ differs by a phase of $\pi$.}
    \label{contour}
\end{figure}

\subsection{The one-loop determinant $\mc{A}$}

In this section, we analyze the one-loop determinant. An imaginary part of the effective action leads to a non-vanishing probability of particle creation, meaning that a determinant of the path fluctuations is essential to the Schwinger effect. As reviewed in section \ref{worldline_formalism}, the determinant of the second variation operator can be diagonalized as
\beq
\det{\mf{H}}=\tilde{\mf{H}}_{ss}\det{\Lambda},
\eeq
where $\Lambda\equiv\mf{H}_{\xi\xi}$ is given by Eq.(\ref{Hess_xixi}). Formally, the determinant of $\Lambda$ is an infinite product of its eigenvalues determined by
\beq
\Lambda\delta\vec{\xi}=\lambda \delta \vec{\xi},
\eeq
and supplied with the boundary conditions
\beq
\delta\vec{\xi}(0)=\delta\vec{\xi}(1)=0.
\eeq
The infinite product is inherently divergent and has to be regularized. The Gelfand-Yaglom method \cite{10.1063/1.1703636, 8cbc8c63-0dd7-3a55-83a4-97485886989f,Levit:1976fv,Kirsten:2004qv,Dunne:2006st} provides just that. The theorem states that
\beq \label{GY-theorem}
\frac{\det{\Lambda}}{\det{\mathbf{\Phi}(1)}}=\frac{\det{\Lambda_0}}{\det{\mathbf{\Phi}_0(1)}},
\eeq
where $\Lambda_0$ is a reference operator, $\mathbf{\Phi}(\tau)$ is the $d\times d$ matrix formed by a set of linearly independent fundamental solutions $\{\vec{u}_i\}$ to the following differential equation and boundary conditions
\beq \bc
\Lambda\vec{u}_i=0\\
\vec{u}_i(0)=0\\
\vec{u}_i'(0)=\vec{w}_i
\ec,\; i=1,2,\cdots, d
\eeq
where $d$ is the rank of $\Lambda$ and $\{\vec{w}_i\}$ is an arbitrary set of $d$ linearly independent vectors\footnote{The construction ensures the regularized determinant is independent of this choice.}.
$\mathbf{\Phi}_0(\tau)$ is defined analogously with respect to the operator $\Lambda_0$. The RHS of Eq.(\ref{GY-theorem}) is absorbed into the definition of the integration measure \cite{Muratore-Ginanneschi:2002sjs, DeWitt:1957at} and thus $\det{\Lambda}\propto \det{\mathbf{\Phi}(1)}$.

The determinants for different $z$ and instanton parameter $\rho_0$ is presented in Fig.(\ref{prefactor}), where $\rho_0$ is translated into the spacetime location $\rho_\times$ where the instanton contributes to the effective action. 

\begin{figure}[htbp]
    \centering
    \includegraphics[width=14cm]{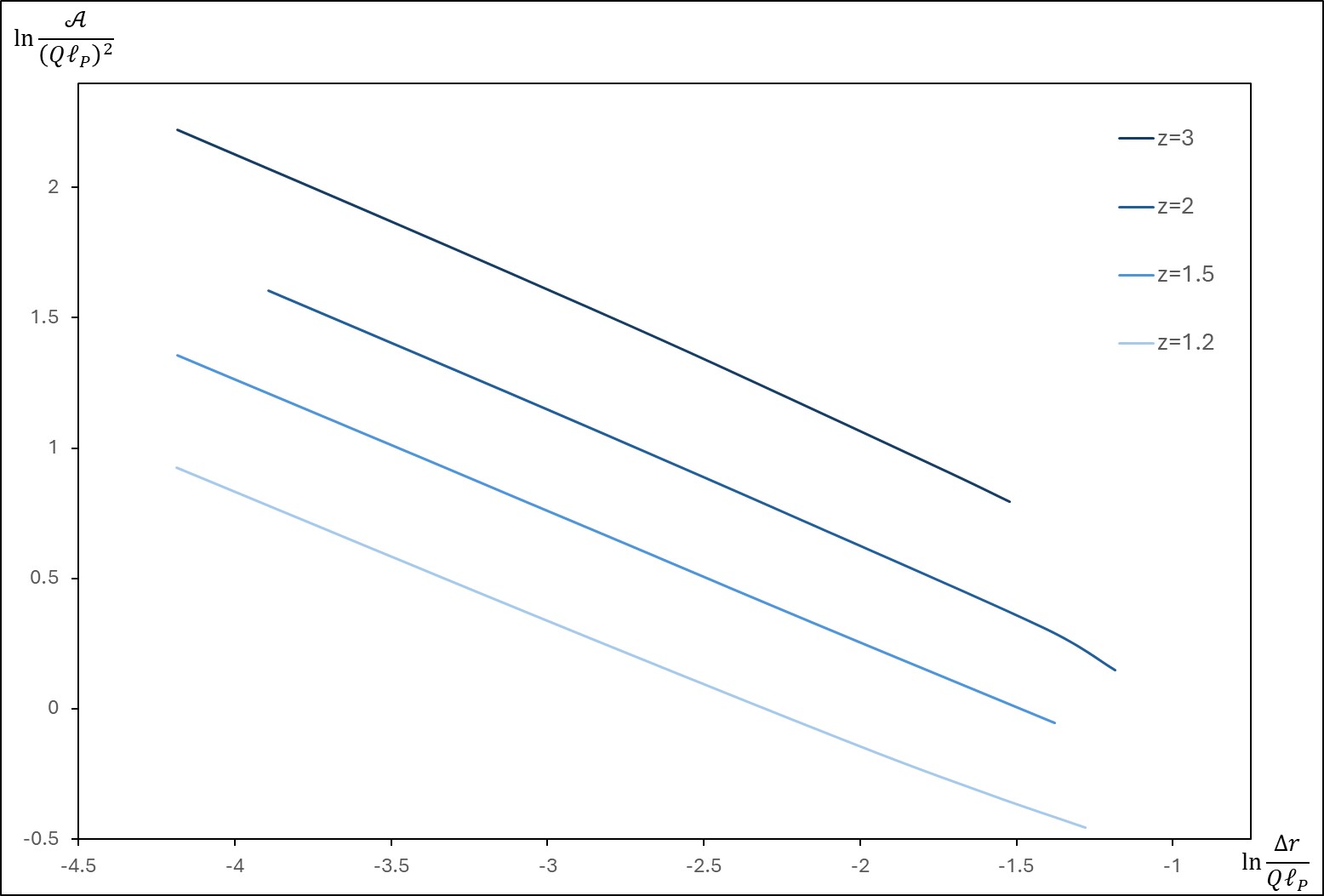}
    \caption{The radial dependence of the prefactor $\mc{A}$. The profile is plotted for some chosen values of $z$. In the range where $\Delta r$ is small, the prefactor and the distance to the horizon satisfy the relation $\mc{A}\propto (\Delta r)^{-1}$.}
    \label{prefactor}
\end{figure}

\subsection{The Schwinger production rate}
The prefactor $\mc{A}$ and exponent $e^{-S_\text{inst}}$ together give the local Schwinger rate and its dependence in the radial coordinate. 
Close to the horizon where the Schwinger effect is dominant, the Schwinger rate is described by the approximate form of
\beq \label{Schwinger_rate}
\Gamma(r)\approx\frac{C\sqrt{z^2-1}}{Q^2\ell_P^2\sqrt{\Delta\tilde{r}}}e^{-(S_{AdS}+B\Delta \tilde{r})},
\eeq
where $\Delta\tilde{r}=\frac{r-r_+}{Q\ell_P}, S_{AdS}=2\pi Qm\ell_P(z-\sqrt{z^2-1})$ and $B=\frac{2\pi Qm\ell_P z(z-1)}{(z^2-1)^\frac{3}{2}}$.
Although the dependence of the prefactor on the charge-to-mass ratio and distance to the horizon is computed numerically, they exhibit general features that can be understood intuitively.

First, the local rate is diverging but integrable when the horizon is approached. The diverging local rate at the horizon is a consequence of restoration of the $AdS_2$ conformal symmetry. Symmetries of this kind lead to zero modes associated with invariance of the stationary action. These zero modes cause the path integral to diverge when summing over the equivalent configurations and contribute an infinite volume factor. In the static RN black hole spacetime, one of such zero modes can be identified as that associated with time translational symmetry, contributing a factorized time interval from the total Schwinger rate. Another zero mode associated with the $AdS_2$ conformal symmetry exists on the horizon. However, in the exterior of the black hole, the conformal symmetry is broken. This means that we expect the zero eigenvalue of the fluctuation operator on the horizon to be continuously uplifted by the separation from the horizon. The uplifted eigenvalue remains small close to the horizon, giving the divergence of the local rate after the path integral. Unlike in pure $AdS_2$ space, however, in the full RN geometry, the volume factor associated with the (near) zero mode of $AdS_2$ cannot be infinite because the $AdS_2$ geometry is a good approximation only up to a finite size. This implies the integrability of the local rate over the black hole exterior. Indeed, the condition for the local rate to be integrable is a natural expectation for particle creation by finite-sized systems such as the black hole. A diverging rate would indicate that the black hole will lose most of its charge in an arbitrarily short amount of time, which is a clear contradiction to the existence of the black hole.

Another feature of the one-loop prefactor is the switch-off behavior near the extremality of the produced particles. Focusing on the near horizon region where the Schwinger effect is dominant, the instanton paths are small deformations from the $AdS_2$ instantons computed in Appendix \ref{AdS-instanton}. The $z$-dependence of the prefactor $\mc{A}\propto \frac{1}{\Bar{s}}$ is captured by the kinetic term of the stationary action $\Bar{s}=\frac{ma}{2}=\frac{\pi Qm\ell_P}{\sqrt{z^2-1}}$. From this we see that $\mc{A}\propto \sqrt{z^2-1}$ and the prefactor continuously approach zero when the particle becomes extremal, switching off the Schwinger effect. The threshold of the Schwinger effect at $z=1$ is reminiscent of the WGC bound. As previously argued from the kinematic viewpoint of preventing naked singularies to form in the backreacted spacetime, particles with $z<1$ cannot cause the decay of extremal black holes in asymptotically flat space. Here the same bound reappears, but seen directly from the dynamics of the charged emission process. While emission of  extremal charged particles by extremal black holes does not lead to exposure of singularities\footnote{The black hole extremality bound receives corrections from higher derivative operators, see \cite{Kats:2006xp,Hamada:2018dde,Loges:2019jzs,Bellazzini:2019xts,Goon:2019faz,Loges:2020trf,Aalsma:2020duv,Cremonini:2021upd,Aalsma:2022knj}, potentially allowing smaller black holes themselves to count towards the superextremal object responsible for the decay of larger black holes. The emission of charged particles by large extremal black holes that we considered is a special case where one of the decay products is microscopic. 
Modular invariance in string theory \cite{Montero:2016tif,Heidenreich:2016aqi} as well as IR consistencies \cite{Andriolo:2018lvp} suggest the existence of a tower of superextremal states. In certain string theoretical setups, the tower was shown to interpolate between extremal black holes and microscopic superextremal particles \cite{Aalsma:2019ryi}.}, such a process is not dynamically favorable. Viewing the Schwinger effect as a quantum mechanical tunneling process driven by the electric force, the aforementioned process sees a cancellation between the electric repulsion and gravitational attraction between the two extremal objects. This effectively diminishes the residue electric force that sources the tunneling, causing a vanishingly small probability for the process to happen. The same threshold can be determined by an entropic reasoning. It is shown in \cite{Parikh:1999mf,Aalsma:2018qwy,Aalsma:2022knj,Aalsma:2023mkz} that the particle creation by black hole is closely related to the change in entropy of the system. Consider emission of one single extremal particle by an extremal black hole. This emission process keeps the black hole on the extremal line but reduces the size of the black hole. The entropy of the emitted extremal particle has to be in the zero momentum state at infinity thus should have zero entropy. Therefore, the emission of extremal particles leads to a net decreased of entropy of the system, suggesting that the process is not entropically favored\footnote{Emission of extremal particles by extremal black holes resembles the $AdS$-fragmentation discussed in \cite{Maldacena:1998uz}, except the particle is not associated with a horizon. The exponential factor that we obtained is consistent with the instanton action in \cite{Maldacena:1998uz} for $AdS$ brane nucleation and with the Brill instanton \cite{Brill:1992ydn}, but our analysis shows a vanishing prefactor which is not evaluated in \cite{Maldacena:1998uz, Brill:1992ydn}. It would be a useful exercise to see whether the vanishing prefactor remains when considering gravitational back-reactions in our analysis or to compute the prefactor associated with the fragmentation process.}.

We note the resemblance of the $z>1$ condition to the charged superradiance condition
\beq
m<\omega <q \Phi_H,
\eeq
where $\Phi_H$ is the electric potential at the horizon and is unity for an extremal RN black hole. In fact, it is not a surprise that the stimulated emission of charged particles, the superradiance effect, is connected to the spontaneous Schwinger effect in such a way. Black holes as quantum systems are thought to obey detailed balancing where the spontaneous and stimulated production imply one another. 

\section{Conclusions} \label{conclusions}
Using the worldline path integral formalism, we compute the spatial profile of the Schwinger production rate in the exterior of an extremal RN black hole. This is the first time a full description of the Schwinger effect outside the RN black hole horizon is given. We notice many interesting aspects of the result that is worth emphasizing, in particular, the characteristic scale of the Schwinger effect outside the black hole, the connection of the Schwinger effect to black hole superradiance and the relation between conditions of non-zero Schwinger rate and bounds on the particle spectrum.

We identify the characteristic scale of the Schwinger effect outside extremal black holes to be the Compton wavelength of the charged particle. From Eq.(\ref{Schwinger_rate}), it is evident that the Schwinger effect is the strongest on the black hole horizon. This is not surprising since the electric field is the strongest there. The scale at which the particle production is not too diminished compared to the horizon rate is determined by the exponential suppression factor. The fall-off scale can be estimated from the value of the exponent. For instance, we define this scale as the distance over which the exponent changes by one, therefore $\frac{B\lambda}{Q\ell_P}\sim 1$. In other words, this scale $\lambda$ is measured by the Compton wavelength
\beq \label{scale}
\lambda\sim c(z)\lambda_c,
\eeq
and $\lambda_c=m^{-1}$ is the Compton wavelength of the particle and $c(z)$ is a $z$-dependent factor\footnote{Concrete values for an electron is given in section \ref{BH_Schwinger_rate}.}.  We note that the Compton length is also the characteristic scale of the superradiant effect of black holes. 
This scale serves as 
an independent check of the detailed balancing principle which relates the spontaneous Schwinger production and the stimulated superradiant scattering effect. The connection can be drawn from both the profile and the strength of the effects. As can be seen from the gravitational atom analysis of superradiance in \cite{Baumann:2019eav}, the extent of a superradiant cloud around the black hole is described by the Compton wavelength, scaled by the gravitational coupling constant, taking the same form as Eq.(\ref{scale}) if we think of $z$ as the relative coupling strength between the electromagnetic and gravitational field felt by the particle. It is also shown that the superradiance rate is controlled by the ratio of the black hole size and the particle's Compton wavelength, giving a highest rate when they are of the same order \cite{Dolan:2007mj,Arvanitaki:2009fg,Dolan:2024qqr}. This suggests an interpolation from suppressed Schwinger effect to catastrophic superradiant instability as we change the parameter $\frac{M \ell_P^2}{\lambda_c}$. The Schwinger effect derived in this paper describes the situation when $\frac{M \ell_P^2}{\lambda_c}\gg 1$. A transition to the unsuppressed regime happens at $\frac{M \ell_P^2}{\lambda_c}\sim O(1)$ when the stationary point approximation breaks down, resembling the tuning of a resonance effect. It is important to note that the results referenced above for the superradiant effect is originally considered for a Kerr black hole, so one has to carefully apply the statements to the charged case. If the results for rotating superradiance generalizes to charged superradiant effect, then there will be a unified picture of the Schwinger effect and the superradiant effect - both are consequences of the instability due to some negative modes induced by the electric field. The spontaneous Schwinger effect describes the decay from the vacuum state while the superradiance phenomenon captures the resonance effect of particle-occupied states with the unstable mode.

The scale of the Schwinger effect being the Compton length should hold more generally for charged black holes embedded in different asymptotic spacetimes because it is the intrinsic scale associated with the particle being created. The scale not only indicates the relevant region for the Schwinger effect, it is also suggestive of the regime where the Schwinger formula starts to break down. Having this in mind, we revisit the scenario that led to the Festina Lente (FL) bound on the particle spectrum in dS space $m^2\gtrsim eH\ell_P^{-1}$. The bound was proposed in \cite{Montero:2019ekk} and was refined in a later paper \cite{Montero:2021otb}. The original idea was to put a constraint on the particle spectrum such that the subsequent evolution of the Nariai spacetime is non-singular. Using the Schwinger formula in dS space derived in \cite{Frob:2014zka}, \cite{Montero:2019ekk} considered the (near) homogeneous rapid discharge of a cosmological-sized charged black hole into light particles throughout the space between the black hole and cosmological horizon. The rapid creation and violent annihilation of the created particles create an oscillating dipole effect that converts the energy of the electric field into radiation, leading to a big crunch of the spacetime. The remedy to this given by the authors in \cite{Montero:2019ekk} was to prevent the rapid discharge by putting a lower bound on the particle mass. We will provide a different physical picture of the Schwinger effect of charged dS black holes in the next paragraph, but before that, it is useful to point out some key ingredients that went into the proposal of the FL bound: (1) a strong Schwinger effect throughout the region between the black hole and cosmological horizons, and (2) the rapid annihilation of charged particles. 

Process (1) causes the black hole to discharge and (2) leads to a collapse of the spacetime, as argued in \cite{Montero:2019ekk}. Our findings strongly motivate one to revisit whether the singular collapse envisioned in \cite{Montero:2019ekk} can happen\footnote{\cite{Aalsma:2023mkz} had also argued that the singular collapse may not occur.}. The coordinate separation\footnote{The coordinate distance sets the $S^2$ size and controls the electric field strength, therefore the spatial profile of the Schwinger production is measured in coordinate distance.} of the black hole and cosmological horizon will be relevant to our discussion and we denote this quantity as $r_u=r_c-r_+$, where $r_c$ is the location of the cosmological horizon and $r_+$ is location of the black hole horizon, see Fig.\ref{RNdS}. When the Compton wavelength is small compared to the horizon separation, $\lambda_c\ll r_u$, particle production happens only near the black hole horizon. Because the charged particles created are highly localized in space, they will not lead to a singular collapse of the full spacetime described in \cite{Montero:2019ekk}. When the separation between the black hole horizon and cosmological horizon is further decreased such that the Compton wavelength becomes larger\footnote{In this limit, the proper distance between the black hole and cosmological horizons stays finite, and is of the dS scale. We note that if the particle's Compton wavelength is even larger than the dS scale, the notion of particle in the spacetime between the horizons breaks down.}, $\lambda_c>r_u$, the charge and mass of the dS black hole will receive significant corrections if the Schwinger rate is unsuppressed. This is because the black hole will form a particle cloud due to the rapid discharge. The particle cloud should have a density distribution that is proportional to the Schwinger rate profile and would carry a significant fraction of the charge and mass of the black hole. If $\lambda_c>r_u$, this cloud extends outside the cosmological horizon, then the actual mass and charge of the black hole appearing in the dS black hole metric will need to be corrected by a potentially large fraction. The significant modification to the mass and charge parameter of a large black hole in dS space (near Nariai black hole) due to the particle cloud calls for a more careful analysis of the particle emission process and the back-reaction on the black hole. This points towards the need of a more complete spatial analysis of the Schwinger production in the general dS black hole spacetime and a proper account of the gravitational back-reaction, which are interesting future extensions of our work.

On the gravitational side, the intuition that the black hole together with the particle cloud of size $\lambda_c$ should be contained within the cosmological horizon has direct implication to the yet open question of whether a near Nariai black hole can dynamically exit the allowed parameter space (the so called shark fin region). When the particle spectrum includes some charged light particles, there will exist a region $\mf{R}$ near the Nariai line in which a black hole should always be considered together with its surrounding charged particle distribution. One should refrain from tracing the evolution of a pure dS black hole starting from $\mf{R}$ because pure black holes cannot be stable here. We speculate that the region $\mf{R}$ modifies the Nariai line and smooths out the evolution of charged dS black holes as they approach $\mf{R}$ from inside the shark fin region, preventing a pathological development of the spacetime.

\begin{figure}[htbp]
    \centering
    \includegraphics[width=14cm]{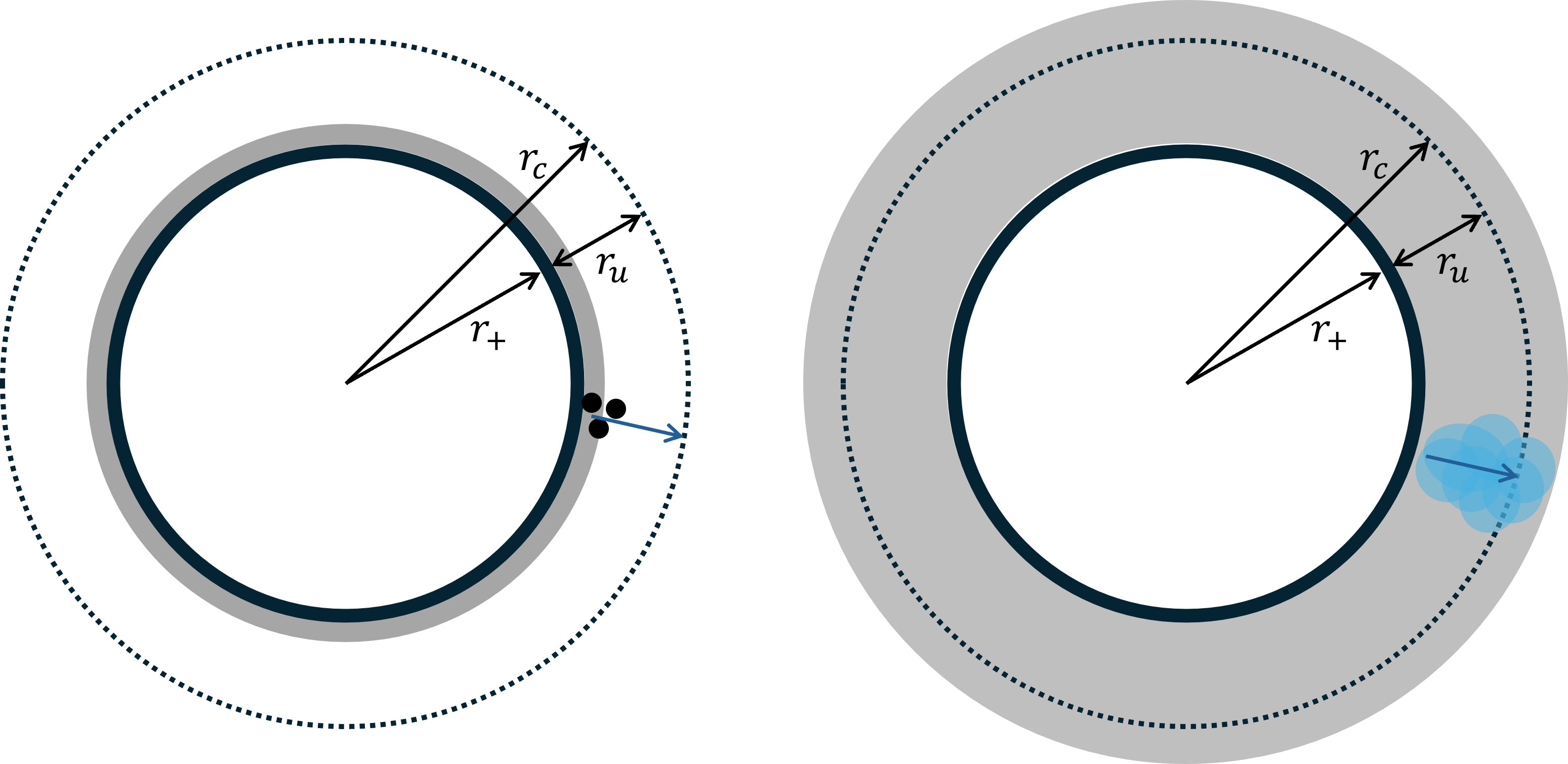}
    \caption{Scales and Schwinger effect of RN black holes in dS. The dotted lines represent the cosmological horizon and the solid lines represent the black hole horizon. The grey shades show the regions with scale $\lambda_c$ where the Schwinger effect is dominant, $\lambda_c\ll r_u$ for the left and $\lambda_c\gtrsim r_u$ for the right. The black dots (left panel) represent the charged particles and the blue blob (right panel) represents the particle cloud, shown only for a particular angular direction. The arrows represent the particle fluxes.}
    \label{RNdS}
\end{figure}

On the particle side, decay of charged dS black holes might provide insights to the WGC. The WGC is a bound on particle spectrum for consistency of quantum gravity allowing for decay of black holes, in a way consistent with the Cosmic Censorship Conjecture. The computation of the Schwinger effect can be indicative of the form of the WGC bound, as seen from the analysis in this paper. It remains an open question of what the Schwinger production rate is between the black hole and cosmological horizons for general dS black holes. This again motivates the application of the worldline approach to the study of Schwinger effect of dS black holes, especially for charged (near) Nariai black holes, to provide hints on the WGC bound in dS space. We will leave this investigation to a future work.

\paragraph{Acknowledgments}
We would like to thank Lars Aalsma, Yoshihiko Abe, Gregory Loges, Miguel Montero, and Jan Pieter van der Schaar for helpful discussions and comments. This work is supported in part by the DOE grant DE-SC0017647.

\appendix
\section{Expansion of the worldline action}
\label{action_expansion}
The worldline action of interest is
\beq
S=\int_0^1d\tau\left(\frac{m^2}{4s} g_{\mu\nu}\Dot{\xi}^\mu\Dot{\xi}^\nu+e A_\mu \dot{\xi}^\mu+s\right).
\eeq
Expanding the kinetic term with respect to the path fluctuation, we obtain
\beq \bal
&g_{\mu\nu}(\xi+\delta\xi)(\Dot{\xi}^\mu+\delta\Dot{\xi}^\mu)(\Dot{\xi}^\nu+\delta\Dot{\xi}^\nu)\\
=&(g_{\mu\nu}+g_{\mu\nu,\sigma}\delta\xi^\sigma+\frac{1}{2}g_{\mu\nu,\sigma\rho}\delta\xi^\sigma\delta\xi^\rho+\cdots)(\Dot{\xi}^\mu+\delta\Dot{\xi}^\mu)(\Dot{\xi}^\nu+\delta\Dot{\xi}^\nu)\\
=&g_{\mu\nu}\Dot{\xi}^\mu\Dot{\xi}^\nu+2g_{\mu\nu}\Dot{\xi}^\mu\delta\Dot{\xi}^\nu+\delta\xi^\sigma g_{\mu\nu,\sigma}\Dot{\xi}^\mu\Dot{\xi}^\nu\\
+&g_{\mu\nu}\delta\Dot{\xi}^\mu\delta\Dot{\xi}^\nu+\delta\xi^\sigma g_{\mu\nu,\sigma}(\delta\Dot{\xi}^\mu\Dot{\xi}^\nu+\Dot{\xi}^\mu\delta\Dot{\xi}^\nu)+\frac{1}{2}\delta\xi^\rho\delta\xi^\sigma g_{\mu\nu,\sigma\rho}\Dot{\xi}^\mu\Dot{\xi}^\nu+\cdots.
\eal \eeq
Similarly, the gauge field term is expanded as
\beq \bal
&A_\mu(\xi+\delta\xi)(\Dot{\xi}^\mu+\delta\Dot{\xi}^\mu)\\
=&(A_\mu+A_{\mu,\nu}\delta\xi^\nu+\frac{1}{2}A_{\mu,\nu\sigma}\delta\xi^\nu\delta\xi^\sigma)(\Dot{\xi}^\mu+\delta\Dot{\xi}^\mu)\\
=&A_\mu\Dot{\xi}^\mu+A_{\mu,\nu}\delta\xi^\nu\Dot{\xi}^\mu+A_\mu\delta\Dot{\xi}^\mu+A_{\mu,\nu}\delta\xi^\nu\delta\Dot{\xi}^\mu+\frac{1}{2}\Dot{\xi}^\mu A_{\mu,\nu\sigma}\delta\xi^\nu\delta\xi^\sigma+\cdots.
\eal \eeq
The action expanded with respect to the path, by expansion order, is
\beq
S^{(0)}=\int_0^1d\tau\left(s+\frac{m^2}{4s} g_{\mu\nu}\Dot{\xi}^\mu\Dot{\xi}^\nu+e A_\mu \dot{\xi}^\mu\right),
\eeq
\beq \bal
S^{(1)}_{\xi}&=\int_0^1d\tau\left[\frac{m^2}{4s}\left(2g_{\mu\nu}\Dot{\xi}^\mu\delta\Dot{\xi}^\nu+\delta\xi^\sigma g_{\mu\nu,\sigma}\Dot{\xi}^\mu\Dot{\xi}^\nu \right)+e\left(A_{\mu,\nu}\delta\xi^\nu\Dot{\xi}^\mu+A_\mu\delta\Dot{\xi}^\mu\right)
\right]\\
&=\int_0^1d\tau\delta\xi^\nu\left\{\frac{m^2}{4s}\left[-2\frac{d}{d\tau}
(g_{\mu\nu}\Dot{\xi}^\mu)+g_{\mu\sigma,\nu}\Dot{\xi}^\mu\Dot{\xi}^\sigma \right]+e\left(A_{\mu,\nu}\Dot{\xi}^\mu-\frac{d}{d\tau}A_\nu\right)
\right\}\\
&=\int_0^1d\tau\delta\xi^\nu\left\{\frac{-m^2}{2s}\left(g_{\mu\nu}\Ddot{\xi}^\mu+g_{\mu\nu,\sigma}\Dot{\xi}^\sigma\Dot{\xi}^\mu-\frac{1}{2}g_{\mu\sigma,\nu}\Dot{\xi}^\mu\Dot{\xi}^\sigma \right)+e\left(A_{\mu,\nu}-A_{\nu,\mu}\right)\Dot{\xi}^\mu
\right\}\\
&=-\int_0^1d\tau\delta\xi^\nu \left\{\frac{m^2}{2s}g_{\mu\nu} \frac{D^2}{d\tau^2}\xi^\mu+eF_{\mu\nu}\Dot{\xi}^\mu\right\}\\
&=-\int_0^1d\tau\delta\xi^\mu \left\{\frac{m^2}{2s}g_{\mu\nu} \frac{D^2}{d\tau^2}-eF_{\mu\nu}\frac{D}{d\tau}\right\}\xi^\nu,
\eal \eeq
\beq \bal
\label{Hess_xixi}
S^{(2)}_{\xi\xi}=&\frac{1}{2}\int_0^1d\tau\delta\xi^\mu\Big\{ \frac{m^2}{2s}\Big[
-g_{\mu\nu}\frac{d^2}{d\tau^2}-\Ddot{\xi}^\sigma g_{\mu\sigma,\nu}-2\Dot{\xi}^\sigma\Gamma_{\sigma\nu\mu}\frac{d}{d\tau}-\Dot{\xi}^\rho\Dot{\xi}^\sigma g_{\mu\sigma,\nu\rho}+\frac{1}{2}\Dot{\xi}^\rho\Dot{\xi}^\sigma g_{\rho\sigma,\mu\nu}\Big]\\
&+e\Big[F_{\mu\nu}\frac{d}{d\tau}+\Dot{\xi}^\sigma(A_{\sigma,\nu\mu}-A_{\mu,\nu\sigma})\Big]\Big\}\delta\xi^\nu,
\eal \eeq
where the following is applied to obtain Eq.(\ref{Hess_xixi})
\beq \bal
&\int_0^1d\tau\left\{
g_{\mu\nu}\delta\Dot{\xi}^\mu\delta\Dot{\xi}^\nu+\delta\xi^\sigma g_{\mu\nu,\sigma}(\delta\Dot{\xi}^\mu\Dot{\xi}^\nu+\Dot{\xi}^\mu\delta\Dot{\xi}^\nu)+\frac{1}{2}\delta\xi^\rho\delta\xi^\sigma g_{\mu\nu,\sigma\rho}\Dot{\xi}^\mu\Dot{\xi}^\nu
\right\}\\
=&\int_0^1d\tau\delta\xi^{\mu}\Big\{-g_{\mu\nu}\delta\Ddot{\xi}^{\nu}-\Dot{\xi}^\sigma g_{\mu\nu,\sigma}\delta\Dot{\xi}^\nu-\left(\Dot{\xi}^\rho\Dot{\xi}^\sigma g_{\mu\sigma,\nu\rho}\delta\xi^\nu+g_{\mu\sigma,\nu}\Ddot{\xi}^\sigma\delta\xi^\nu+g_{\mu\sigma,\nu}\Dot{\xi}^\sigma\delta\Dot{\xi}^\nu \right)\\
&+\Dot{\xi}^\sigma g_{\sigma\nu,\mu}\delta\Dot{\xi}^\nu
+\frac{1}{2}\Dot{\xi}^\rho\Dot{\xi}^\sigma g_{\rho\sigma,\mu\nu}\delta\xi^\nu \Big\}\\
=&\int_0^1d\tau\delta\xi^\mu\Big\{
-g_{\mu\nu}\frac{d^2}{d\tau^2}-g_{\mu\sigma,\nu}\Ddot{\xi}^\sigma-2\Dot{\xi}^\sigma\Gamma_{\sigma\nu\mu}\frac{d}{d\tau}-\Dot{\xi}^\rho\Dot{\xi}^\sigma g_{\mu\sigma,\nu\rho}+\frac{1}{2}\Dot{\xi}^\rho\Dot{\xi}^\sigma g_{\rho\sigma,\mu\nu}\Big\}\delta\xi^\nu,
\eal \eeq
\beq \bal
&\int_0^1d\tau\left\{
A_{\mu,\nu}\delta\xi^\nu\delta\Dot{\xi}^\mu+\frac{1}{2}\Dot{\xi}^\mu A_{\mu,\nu\sigma}\delta\xi^\nu\delta\xi^\sigma \right\}\\
=&\int_0^1d\tau\left\{
\frac{1}{2}A_{\mu,\nu}\delta\xi^\nu\delta\Dot{\xi}^\mu+\frac{1}{2}A_{\nu,\mu}\delta\xi^\mu\delta\Dot{\xi}^\nu+\frac{1}{2}\Dot{\xi}^\sigma A_{\sigma,\nu\mu}\delta\xi^\nu\delta\xi^\mu \right\}\\
=&\frac{1}{2}\int_0^1d\tau\delta\xi^\mu\left\{
A_{\nu,\mu}\delta\Dot{\xi}^\nu-A_{\mu,\nu}\delta\Dot{\xi}^\nu-A_{\mu,\nu\sigma}\Dot{\xi}^\sigma\delta\xi^\nu+\Dot{\xi}^\sigma A_{\sigma,\nu\mu}\delta\xi^\nu\delta \right\}\\
=&\frac{1}{2}\int_0^1d\tau\delta\xi^\mu\left\{F_{\mu\nu}\frac{d}{d\tau}+\Dot{\xi}^\sigma(A_{\sigma,\nu\mu}-A_{\mu,\nu\sigma})\right\}\delta\xi^\nu.
\eal \eeq
The expansion with respect to the proper time can be easily computed since it is not dynamical. The results are
\beq \bc
S^{(1)}_s=\int_0^1d\tau (1-\frac{m^2}{4s^2}g_{\mu\nu}\Dot{\xi}^\mu\Dot{\xi}^\nu)\delta s\\
S^{(2)}_{ss}=\int_0^1d\tau \left(\frac{m^2}{2s^3}g_{\mu\nu}\Dot{\xi}^\mu\Dot{\xi}^\nu\right) \delta s^2\\
S^{(2)}_{s\xi}=\int_0^1d\tau\delta s \left(\frac{m^2}{2s^2}g_{\mu\nu} \frac{D^2}{d\tau^2}\xi^\nu\right) \delta\xi^\mu
\ec \eeq

\section{Green's function of matrix differential operators}
\label{Greens_function}
Consider the matrix generalization of the Sturm-Liouville operator defined on $\tau\in[0,1]$
\beq
L=\frac{d}{d\tau}P\frac{d}{d\tau}+Q,
\eeq
where $P,Q$ are matrices. For what is pertained to this paper, the boundary condition is Dirichlet. Assuming that the kernel of the operator is trivial, the inverse is defined as the solution of
\beq
LG(\tau,\tau')=\delta(\tau-\tau'),
\eeq
which can be constructed as a gluing of two solutions
\beq
G(\tau,\tau')=\Theta(\tau'-\tau)Y_L(\tau)A(\tau')+\Theta(\tau-\tau')Y_R(\tau)B(\tau'),
\eeq
where $Y_{L,R}$ are matrices formed by independent solutions to $L\vec{y}_{L,R}=0$ satisfying boundary conditions $\vec{y}_L(0)=\vec{y}_R(1)=0$ and $A,B$ are matrix coefficients.
Continuity of $G$ and the jump of derivative due to the delta function lead to the condition
\beq \bc
Y'_L(\tau')A(\tau')-Y'_R(\tau')B(\tau')=-P^{-1}\\
Y_L(\tau')A(\tau')-Y_R(\tau')B(\tau')=0
\ec. \eeq
Solving for $A,B$, 
\beq \bc
A=\left[P (Y'_R Y_R^{-1} Y_L-Y'_L)\right]^{-1}\\
B=\left[P(Y'_R-Y'_LY_L^{-1}Y_R)\right]^{-1}
\ec, \eeq
the full Green's function is then
\beq \bal
G(\tau,\tau')&=\Theta(\tau'-\tau)Y_L(\tau)\left[P (Y'_R Y_R^{-1} Y_L-Y'_L)\right]^{-1}(\tau')\\
&+\Theta(\tau-\tau')Y_R(\tau)\left[P(Y'_R-Y'_LY_L^{-1}Y_R)\right]^{-1}(\tau').
\eal \eeq

When the operator $L$ has zero eigenvalues, a pseudo-inverse can be defined by projecting onto the orthogonal space. Making use of the eigenvalue representation of the Green's function, we can construct
\beq
G'(\tau,\tau')=\lim_{\epsilon\rightarrow0}\left(G_\epsilon(\tau,\tau')-\sum_m\frac{y_{0m}^T(\tau)y_{0m}(\tau')}{\epsilon}\right),
\eeq
where $G_\epsilon$ is the Green's function of the shifted operator $L_\epsilon=L+\epsilon$ and $y_{0m}$ is the zero mode of $L$.

\section{Worldline instantons in $AdS_2$ space}
\label{AdS-instanton}
In this appendix, we will compute the instanton path and action in the Poincar\'e and global coordinates of $AdS_2$. The Poincar\'e coordinate (in Euclidean signature) is
\beq
ds^2=L^2\frac{dt^2+dr^2}{r^2}.
\eeq
The gauge field is taken to be $A=\frac{EL^2}{r}dt$, corresponding to a constant electric field in $AdS_2$. The equations of motions is
\beq \label{geodesic_Poincare1} \bc 
\Dot{r}=\pm \frac{ar}{L}\sqrt{1-z^2(1-\tilde{\omega} r)^2}\\
\Dot{t}=\frac{azr}{L}(\tilde{\omega} r-1)
\ec, \eeq
where $z=\frac{eEL}{m}, \tilde{\omega}=\frac{\omega}{eEL^2}$. Solutions to the above equation are circles. This can be seen after substituting $\Dot{t}=\Dot{r}\frac{dt}{dr}$ into the second line of Eq.(\ref{geodesic_Poincare1}),
\beq \label{geodesic_Poincare2}
\frac{dt}{dr}=\pm\frac{z(\tilde{\omega} r-1)}{\sqrt{1-z^2(1-\tilde{\omega} r)^2}},
\eeq
which describes circles that are tangent to $t=\pm \frac{r}{z^2-1}$. We can thus parametrize the solutions by
\beq \label{AdS_parametrize} \bc
t=-\frac{1}{\tilde{\omega} z}\sin{\theta(\tau)}\\
r=\frac{1}{\tilde{\omega}}-\frac{1}{\tilde{\omega} z}\cos{\theta(\tau)}
\ec. \eeq
The concrete form of $\theta(\tau)$ is obtained by inserting Eq.(\ref{AdS_parametrize}) back to Eq.(\ref{geodesic_Poincare1})
\beq
\theta(\tau)=2\arctan{\left[\sqrt{\frac{z-1}{z+1}}\tan{\left(\frac{a\tau}{2L}\sqrt{z^2-1} \right)} \right]}.
\eeq
The periodic condition of the instanton path sets $a=\frac{2\pi L}{\sqrt{z^2-1}}$ and the Poincar\'e coordinates restricts the valid paths to the $r>0$ region, requiring $z>1$.

The action can then be computed by reading off the kinetic contribution
\beq \label{AdS_inst_k}
ma=\frac{2\pi mL}{\sqrt{z^2-1}}
\eeq
and computing the gauge field contribution
\beq \bal
S_A&=e\int_0^1 d\tau A_0 \Dot{t}=e\int_0^1d\tau\frac{EL^2}{r}\frac{azr}{L}(\tilde{\omega}r-1)\\
&=2zm\int_{r_1}^{r_2}\frac{z(\tilde{\omega}r-1)}{r\sqrt{1-z^2(1-\tilde{\omega}r)^2}}dr\\
&=2\pi mL(z-\frac{z^2}{\sqrt{z^2-1}}).
\eal \eeq
The full instanton action is therefore
\beq
S=2\pi mL(z-\sqrt{z^2-1}),
\eeq
in agreement with that in \cite{Pioline:2005pf}.

In global coordinates\footnote{The same symbol $r$ is used here, which should not be confused with the radial coordinate in the Poincar\'e patch.},
the metric is
\beq
ds^2=(1+\frac{r^2}{L^2})dt^2+\frac{dr^2}{1+\frac{r^2}{L^2}}
\eeq
and the gauge field $A=Erdt$. The corresponding geodesic equation is
\beq \label{geodesic_global} \bc
\Dot{r}=\pm a\sqrt{(1+\frac{r^2}{L^2})-\frac{(eEr-\omega)^2}{m^2}}\\
\Dot{t}=\frac{a}{m}\frac{\omega-eEr}{1+\frac{r^2}{L^2}}
\ec. \eeq
After defining the dimensionless coordinate$\rho=\frac{r}{L}$ and parameters $\tilde{E}=EL^2, \tilde{\omega}=\omega L, z^2=\frac{e^2\tilde{E}^2}{m^2L^2}, \rho_0=\frac{\tilde{\omega}}{e\tilde{E}}$, we rewrite the equation as
\beq \bal \label{AdS_radial}
\Dot{\rho}&=\pm \frac{a}{L}\sqrt{(1+\rho^2)-z^2(\rho-\rho_0)^2}\\
&=\pm\frac{a}{L}\sqrt{z^2-1}\sqrt{\frac{1-z^2\rho_0^2}{z^2-1}+\frac{2z^2\rho_0}{z^2-1}\rho-\rho^2}\\
&=\pm\frac{a}{L}\sqrt{z^2-1}\sqrt{b^2-(\rho-\Bar{\rho})^2},
\eal \eeq
where we further defined $b^2=\frac{z^2-1+z^2\rho_0^2}{(z^2-1)^2}>0$ and $\Bar{\rho}=\frac{z^2}{z^2-1}\rho_0$. For Eq.(\ref{AdS_radial}) to yield periodic solutions, it is required that $z>1$. The above equation is integrated to obtain
\beq
\arcsin{\frac{\rho-\Bar{\rho}}{b}}=\frac{a}{L}\sqrt{z^2-1}(\tau-\tau_0),
\eeq
and the instanton path is\footnote{Solutions in the global coordinates are not circular but can be mapped to the circles in Poincar\'e coordinates derived in the earlier part of this appendix and reported in \cite{Pioline:2005pf, COMTET1987185}.}
\beq \label{AdS_sol}
\rho=\Bar{\rho}+b\sin{[\frac{a}{L}\sqrt{z^2-1}(\tau-\tau_0)]}.
\eeq
Periodicity translates to $a=\frac{2\pi L}{\sqrt{z^2-1}}$. To check that the time coordinate is also periodic, we eliminate proper time from Eq.(\ref{geodesic_global}) using $\Dot{t}=\Dot{\rho}\frac{dt}{d\rho}$
\beq \label{t-r-relation}
\frac{dt}{dr}=\frac{Lz(\rho_0-\rho)}{(1+\rho^2)\sqrt{1+\rho^2-z^2(\rho-\rho_0)}}.
\eeq
We will not show the integration here but it can be computed using a similar complex contour integral performed in the following to compute the instanton action. It is not hard to checked that integration of Eq.(\ref{t-r-relation}) between the two turn-around points of $\rho$ is zero, assuring that the paths are closed.

We next compute the action using the instanton solution Eq.(\ref{AdS_sol})
\beq \bal
S_A&=e\int_0^1 d\tau A_0\Dot{t} = e\int d\tau\frac{A_0(\omega-eA_0)}{\tilde{m}g_{00}}\\
&=\frac{ea}{mL^2}\int_0^1 \frac{d\tau}{1+\rho^2}\tilde{E}\rho(\tilde{\omega}-e\tilde{E}\rho)\\
&=\frac{2e^2\tilde{E}^2}{mL\sqrt{z^2-1}}\int_{\rho_-}^{\rho_+} \frac{d\rho}{1+\rho^2}\frac{\rho(\rho_0-\rho)}{\sqrt{b^2-(\rho-\Bar{\rho})^2}},
\eal \eeq
where we have changed the integration variable from $d\tau$ to $d\rho$ in the third line. The original integral is along a periodic curve parametrized by $\tau\in[0,1]$, so upon switching to $d\rho$, the path is represented by two oppositely orientated paths $\rho=\rho_-\rightarrow \rho_+$ and $\rho=\rho_+\rightarrow \rho_-$ with Jacobians of opposite signs.

To evaluate the integral, we write $\frac{\rho(\rho_0-\rho)}{1+\rho^2}=-1+\frac{1+\rho_0\rho}{1+\rho^2}$, separating it into two parts
\beq \bc
I_1=\int_{\rho_-}^{\rho_+} \frac{d\rho}{\sqrt{(\rho_+-\rho)(\rho-\rho_-)}}\\
I_2=\int_{\rho_-}^{\rho_+} \frac{d\rho}{1+\rho^2}\frac{1+\rho_0\rho}{\sqrt{(\rho_+-\rho)(\rho-\rho_-)}}
\ec. \eeq
The first integral is just $I_1=\pi$ and the second can be computed via the residue theorem by taking a contour very similar to that in Fig.\ref{contour}. The result is
\beq
I_2=\frac{\pi}{[(\rho_+^2+1)(\rho_1^2+1)]^\frac{1}{4}}[\sin{(\frac{\varphi_++\varphi_-}{2})}-\rho_0\cos{(\frac{\varphi_++\varphi_-}{2})}]
\eeq
where $\varphi_\pm=\arg(-\rho_\pm+i)$. Using the root relations of $\rho_\pm$ from Eq.(\ref{AdS_radial}), the denominator is computed as $z\sqrt{\frac{1+\rho_0^2}{z^2-1}}$ and the terms in the braket reduce to $\sqrt{1+\rho_0^2}$.
The total action can finally be expressed as
\beq \bal
S&=2\pi mL[\frac{1}{\sqrt{z^2-1}}+\frac{z^2}{\sqrt{z^2-1}} (\frac{\sqrt{z^2-1}}{z}-1)]\\
&=2\pi mL(z-\sqrt{z^2-1}),
\eal \eeq
where $z=\frac{eEL^2}{mL}=\frac{eEL}{m}$.
This is exactly the same result obtained in the Poincar\'e patch and recovers the flat space instanton action for large $L$, $\lim_{L\rightarrow \infty}S=\frac{\pi m^2}{eE}$.

\bibliographystyle{unsrt}
\bibliography{references}

\end{document}